\documentclass[aps,twocolumn, longbibliography,
 amsmath,10pt, prl,
]{revtex4-2}
\usepackage[T1]{fontenc}
\usepackage{lmodern}
\usepackage{booktabs}
\usepackage{xcolor}
\usepackage{graphicx,subcaption}\usepackage{dcolumn}\usepackage{bm}\usepackage[colorlinks,citecolor=blue,linkcolor=blue,hyperindex]{hyperref}
\usepackage{babel}
\usepackage{comment}
\usepackage{soul}
\usepackage{mathtools}
\usepackage{amssymb}

\usepackage{tikz,pgfplots}
\usepackage{tikz-3dplot}
\pgfplotsset{compat=1.18}
\usetikzlibrary{calc}
\usetikzlibrary{arrows.meta,bending,positioning,fit,backgrounds}
\usetikzlibrary{decorations.markings}
\usepackage{standalone}

\usepackage[mathscr]{euscript}
\usepackage{etoolbox}
\usepackage{cleveref}
\apptocmd{\sloppy}{\hbadness 10000\relax}{}{}
\newcommand{\panelref}[2]{\hyperref[#1]{\getrefnumber{#1}#2}}

\newcommand{\quu}{\textbf{q}}

\newcommand{\Dee}{\mathcal{D}_{\theta}}
\newcommand{\Dpr}{\mathcal{D}_{\theta'}}
\newcommand{\tvphi}{\tilde{\varphi}}

\newcommand{\tmop}[1]{\ensuremath{\operatorname{#1}}}

\begin{document}

\title{Critical theory of Pomeranchuk transitions via high-dimensional bosonization}

\author{Zhengfei Hu}
\thanks{These authors contributed equally to this work.}

\author{Jaychandran Padayasi}
\thanks{These authors contributed equally to this work.}
\affiliation{Department of Physics and National High Magnetic Field Laboratory,
Florida State University, Tallahassee, Florida 32306, USA}

\author{O\u{g}uz T\"urker}
\affiliation{Department of Physics and National High Magnetic Field Laboratory,
Florida State University, Tallahassee, Florida 32306, USA}

\author{Kun Yang}
\affiliation{Department of Physics and National High Magnetic Field Laboratory,
Florida State University, Tallahassee, Florida 32306, USA}

\date{\today}

\begin{abstract}
We use high-dimensional bosonization to derive an effective field theory that describes the Pomeranchuck transition in isotropic two-dimensional Fermi liquids. We find that the transition is triggered by the softening of an eigenmode that leads to spontaneous Fermi surface distortion. The resultant theory in terms of this critical mode has dynamical critical exponent $z = 2$ and the upper critical dimension is $d_c = 4-z= 2$. As a result the system is at the upper critical dimension in 2D, resulting in a Gaussian fixed point with a marginally irrelevant quartic perturbation. 
\end{abstract}

\maketitle

\textit{Introduction.} Quantum phase transitions in itinerant electronic systems have been one of the key drivers of research in condensed matter physics for many years \cite{Sachdev_2011}. Among them the simplest are Pomeranchuk transitions that result in spontaneous deformation of the Fermi surface, as Landau-Fermi liquid theory \cite{PinesNozieres,GiulianiVignale,GirvinYang} already allows for them (without additional ingredients), and predicts the exact values of Landau parameters that trigger these transitions. In this Letter, we study the critical properties of Pomeranchuk transitions in an isotropic two-dimensional (2D) Fermi liquid, where rotation symmetry is broken spontaneously when a Landau parameter $f_\ell$ reaches $-1$ \cite{PinesNozieres,GirvinYang}. Such transitions are closely related to the physics of nematic Fermi liquid phases that have been observed experimentally in $\text{Sr}_3\text{Ru}_2\text{O}_7$ (see \cite{ExperimentsReview} for a review) and iron-based superconductors \cite{Fernandes2016}.

A Ginzburg-Landau-Wilson type theory for the Pomeranchuk instabilities (PI) was first formulated in Ref. \cite{Oganesyan2001} using the framework of Hertz-Millis theory \cite{Hertz1976, Millis1993}. In this approach, four-fermion interactions are decoupled by Hubbard–Stratonovich transformations. The fermionic degrees of freedom are then integrated out, resulting in an effective theory expressed in terms of the bosonic Hubbard–Stratonovich field which plays the role of order parameter. Ref. \cite{Oganesyan2001} finds that the critical theory is described by a Landau-damped mode with dynamical exponent $z = 3$ and upper critical dimension $d_c = 4-z= 1$, which is a common feature of Hertz-Millis type theories, leading to mean-field critical properties in all dimensions where such an instability can occur. Subsequent studies \cite{MaslovChubukov2010PRB,Metlitski2010} have suggested that this approach might be problematic, as integrating out the gapless fermions is widely believed to be an uncontrolled procedure. Numerical studies on the closely related Ising nematic transition \cite{Schattner2016, Berg2019} and ferromagnetic transition \cite{Liu2022} using sign-problem free quantum Monte Carlo see regimes where the susceptibility follows $z = 2$ scaling which is inconsistent with the corresponding Hertz-Millis type theory.

In this Letter, we approach the PI using high-dimensional bosonization (HDB) \cite{Luther:1979,Haldane:1994,Houghton1992,houghton1994,Houghton2000,CastronetoFradkin1994PRL,CastronetoFradkin1994,Khveshchenko_1994,Khveshchenko_1995,Kopietz_1997,Delacretaz2022, Chen2025} which avoids the issues encountered in the Hertz-Millis treatment. 
In one-dimensional (1D) systems, bosonization is a powerful yet simple tool to study low energy physics due to the isomorphism between bosonic Fock space and fixed-charge sector of fermionic Fock space (see Chapter 14 of \cite{Kac_1990}).
In particular, one of us \cite{Kun2004} used it to study ferromagnetic transition in 1D metals. Its higher dimensional counterpart has successfully reproduced the results of Fermi liquid theory, but yielded relatively few new results thus far \cite{Ding:2012}. We believe it is particularly suitable to study Pomeranchuk transitions, because they are driven by Fermi liquid interactions which are incorporated {\em exactly} by bosonization (instead of perturbatively in Hertz-Millis theory). In fact, one of us \cite{Kun2005} made this suggestion 20 years ago. Although that premature attempt is now superseded by the present work (which builds on insights provided by many other works since then as we comment on below), it did anticipate some of our key results here.

In the following we first reformulate HDB in 2D in a way that is not only most convenient for our purpose, but also closest to the much more familiar 1D bosonization in its final form. We then analyze the quadratic part of the bosonized action close to the instability, and show that PI is triggered by softening of an {\em exact} eigenmode. While this critical mode is damped in the (symmetric) Fermi liquid phase, it becomes increasingly under-damped upon approaching PI, where damping disappears. This is the biggest difference between our approach and that of Ref. \cite{Oganesyan2001} and Hertz-Millis type of theories, and leads to dynamical exponent $z=2$ instead of 3 as anticipated in \cite{Kun2005}. The resultant Gaussian fixed point describing the critical point is at its upper critical dimension $d_c = 4-z= 2$, which we analyze toward the end. We put special emphasis on $\ell=2$ channel physics, while generalizations to other channels are straightforward.

\begin{figure}[hbt]
  \centering
  \resizebox{\linewidth}{!}{
    \includegraphics{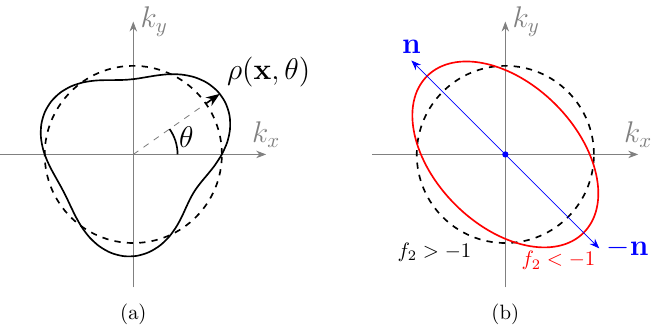}
  }
  \caption{Fluctuations and instabilities of the Fermi surface. (a) Fields $\rho(\mathbf{x}, \theta)$ are local fluctuations of the Fermi surface (solid line) from the circular reference Fermi surface (dashed line) that corresponds to the isotropic ground state. (b) Pomeranchuk instability via condensation of $\rho$ in the $\ell = 2$ channel, resulting in a nematic state with an elongated Fermi surface (red solid line). $\mathbf{n}$ is a unit vector indicating the orientation of the nematic Fermi surface, which is equivalent to $-\mathbf{n}$.}
  \label{fig:fermisurfaces}
\end{figure}

\textit{Gaussian action.} Our starting point is the bosonized action describing 2D isotropic spinless Fermi liquid written in terms of the field $\phi(t,\mathbf{x},\theta)$ whose normal derivative $\rho(t,\mathbf{x}, \theta) \equiv \mathbf{n}_\theta\cdot\nabla\phi,\mathbf{n}_\theta=(\cos \theta,\sin \theta)$ describes local flucations of Fermi surface; see Fig. \panelref{fig:fermisurfaces}{a}. Its quadratic part $S_0[\phi]$ is given by
\begin{multline}
\label{eq:GaussianAction}
  S_{0} = \int dt\ d^2x \Big\{\int^{2\pi}_{0}\frac{d\theta}{2\pi} \mathbf{n}_\theta \cdot \nabla\phi\left(\dot{\phi} - v_F \mathbf{n}_\theta\cdot \nabla \phi\right) \\
  - v_F \int^{2\pi}_{0} \frac{d\theta d\theta'}{(2\pi)^2} F(\theta - \theta')\ \mathbf{n}_\theta\cdot\nabla\phi(\theta)\ \mathbf{n}_{\theta'}\cdot \nabla\phi(\theta')\Big\}
\end{multline}
where $F(\theta - \theta')$ is the Fermi-liquid interaction and $v_F$ is the Fermi
velocity.  To the best of our knowledge, it was first put in the present form in
Ref. \cite{Khveshchenko_1995} (inspired by earlier works
e.g. \cite{Haldane:1994}, see also
\cite{houghton1994,CastronetoFradkin1994,Houghton2000, Delacretaz2022}; similar
construction was also discussed in the context of quantum hall droplet
\cite{Karabali_2004a,Karabali_2004b}).  It is derived from the
fermion effective action for isotropic spinless Fermi liquid fixed point
\cite{shankar1994,Chitov_1995,Metzner_1998,Houghton2000}, where all other interaction terms
are irrelevant. 
Variable $\theta$ is the momentum space ``patch index'' in old literature \cite{Houghton2000,CastronetoFradkin1994,Kopietz_1997}. Such continuous treatment brings technical convenience and is harmless in the large patch number limit.

In 1D bosonization, the density and current are typically separated by defining $(\partial_x\phi, \partial_x\theta) = \rho_R \pm \rho_L$ where $\rho_R\ (\rho_L)$ is the density of right (left) movers \footnote{$\rho_R$ and $\rho_L$ are actually the 1D counterpart of $\rho(\theta)$ in 2D, as there are only two directions in 1D, while there is a continuum of them in 2D parameterized by $\theta$.}. Similar to the standard manipulations there \cite{Giamarchi2003Book}, we separate the odd and even angular harmonics of the Fermi surface label $\theta$ by writing
$\varphi(\theta) = (\phi(\theta) + \phi(\theta+ \pi))/2, \tvphi(\theta) =
(\phi(\theta) - \phi(\theta+ \pi))/2$:
\begin{multline}
  \label{eq:varphiAction}
  S_{0} = \int dt\ d^2x  \Big\{\int_0^\pi \frac{d\theta}{\pi} \big[ 2\dot{\tvphi}\Dee\varphi -  v_F(\Dee\varphi)^2 -  v_F(\Dee\tvphi)^2 \big] \\  
   - v_F \int_0^\pi \frac{d\theta d\theta'}{\pi^2} [\tilde{f}(\theta - \theta') \Dee\varphi\Dpr\varphi 
    + f(\theta-\theta')\Dee\tvphi\Dpr\tvphi] \Big\}
\end{multline}
where we use the notation
$\mathbf{n}_\theta\cdot\nabla \equiv \Dee $\footnote{$\Dee$ is odd under
  $\theta \rightarrow \theta + \pi$}. In the above action we also define the
even $f \equiv (F(\theta) + F(\theta +\pi))/2$ and odd
$\tilde{f} \equiv (F(\theta) - F(\theta +\pi))/2$ parts of the Fermi-liquid
interaction.
To write down a configuration space version of the phase space action Eq.~\eqref{eq:varphiAction} \setcounter{footnote}{10}\footnote{Fields $\varphi$ and $\tvphi$ are not
  independent in the sense that $(\varphi, \Pi_{\varphi} \sim \Dee \tvphi)$ and
  $(\tvphi, \Pi_{\tvphi} \sim \Dee \varphi)$ are merely two different conjugate pairs to
  parametrize the phase space of fields. Either parametrization admits a
  polarization and allows a configuration space form of the action
  Eq. \eqref{eq:varphiAction}. Here we choose
  $(\tvphi, \Pi_{\tvphi} \sim \Dee \varphi)$ since we want to study even channel
  densities. For odd channel densities, one should do the other way around.}, we note that
$\rho(\theta) = \Dee (\varphi(\theta)+\tvphi(\theta))$ and Fourier transform Eq.~\eqref{eq:varphiAction}
 with respect to both $\textbf{x}$ and $\theta$:

\begin{multline}
  S_{0} = \int \frac{dt d^2q}{(2\pi)^2} \Big\{- \sum_{\ell \in \text{odd}} v_F \tilde{g}_\ell \rho_{-\ell}(-\quu) \left( \rho_{\ell}(\quu) - \frac{\dot{\tvphi}_\ell (\quu) }{v_F \tilde{g}_\ell} \right) \\
  - \sum_{\ell,\ell' \in \text{odd}} v_F|\quu|^2 \tvphi_{-\ell}(-\quu)M_{\ell\ell'}(\theta_\quu,\{g\})\tvphi_{\ell'}(\quu) \Big\}.
  \label{eq:action_1}
\end{multline}
Here, Fourier transform with respect to $\theta$ is defined as $\xi_\ell = \int_0^{2\pi} \frac{d\theta}{2\pi} e^{i \ell \theta} \xi(\theta) $ for general $\xi$,
$(g_\ell, \tilde{g}_\ell) \equiv(1 + f_\ell, 1+\tilde{f}_\ell)$ and the matrix
$M_{\ell \ell'}$ (see Eq.~\eqref{eq:coupling-mat-0}) is a tridiagonal matrix that
depends only on the even channel parameters $g_\ell$.  Integrating out $\rho$, we get the effective action in terms of $\tvphi$ as
\begin{multline}
  S_{0}[\tvphi] = \int \frac{dt d^2q}{(2\pi)^2} \Big\{ \sum_{\ell \in \text{odd}} \frac{\dot{\tvphi}_{-\ell}(-\quu)\dot{\tvphi}_\ell(\quu)}{v_F \tilde{g}_\ell }
  \\ - \sum_{\ell,\ell' \in \text{odd}} v_F|\quu|^2 \tvphi_{-\ell}(-\quu)M_{\ell\ell'}(\theta_\quu,\{g\})\tvphi_{\ell'}(\quu) \Big\}.
  \label{eq:S_0}
\end{multline}
The coupling matrix $M_{\ell\ell'}$ depends on $\theta_{\quu}$ since
$\tvphi_{\ell}(\quu)$ carries angular momentum $-\ell$ (see below). 
In the bosonized action, rotation symmetry manifests itself as the
simultaneous rotation of space orientation \textit{and} the Fermi surface label
$\theta$ by the same angle
$(R_{- \alpha} .  \phi) (\mathbf{x}, \theta) \equiv \phi (R_{\alpha} .\mathbf{x}, \theta + \alpha)$
\footnote{Symmetry group is the diagonal subgroup of
  $\tmop{SO} (2) \times U (1)$.}, which results in
$(R_{- \alpha} .  \tvphi_{\ell}) (\mathbf{x}) = e^{i \ell \alpha} \tvphi_{\ell} (R_{\alpha}
.\mathbf{x})$.
$M_{\ell\ell'}$ can be made a constant if we dress $\tvphi_\ell$ by a phase of angular momentum $\ell$. Defining
$\tvphi_{\ell} (\quu) = - ie^{- i \ell \theta_{\quu}} \tvphi'_{\ell} (\quu)$, we find that
$\tvphi'_{\ell}$ behaves trivially under spatial rotation
$(R_{- \alpha} . \tvphi'_{\ell}) (\quu) = \tvphi'_{\ell} (R_{\alpha} .\quu)$. Rewriting the
action in terms of $\tvphi'$ renders the coupling $M'_{\ell\ell'}$ constant (see
Eq.~\eqref{eq:coupling-mat}), and allows the action to be expressed in a simple
real space form
\begin{equation}
  S_0 = \int dt d^2x \sum_{\ell, \ell' \in \text{odd}}
  \Big\{\delta_{\ell\ell'} \frac{\dot{\tvphi}'_{-\ell}\dot{\tvphi}'_{\ell'}}{v_F\tilde{g}_\ell}
  - v_F \nabla \tvphi'_{-\ell} M'_{\ell\ell'}(\{g\}) \nabla \tvphi'_{\ell'}\Big\}.
  \label{eq:S_1}
\end{equation}
We remark here that all the manipulations so far are, in spirit, parallel to 1D bosonization, although expressions like Eq.~\eqref{eq:S_0} and Eq.~\eqref{eq:S_1} were never obtained for higher dimensions before. In 1D, one has the freedom to keep either the bosonic field $\phi(x)$ or its dual $\theta(x)$ \cite{Giamarchi2003Book}, similarly here we have the freedom to keep
$\varphi_{\ell}(\mathbf{x})$ or $\tvphi_{\ell}(\mathbf{x})$, which are dual to each other \cite{Note11}. On the other hand, in 1D the Gaussian theory out of (abelian) bosonization is a single free boson, while here we have
a family of coupled bosons, one for each angular momentum channel (either all even or all odd).

\textit{The Pomeranchuk instability.}
Note that the coefficient $\tilde{g}_\ell^{-1}$ in the dynamical term of Eq. \eqref{eq:S_1} already signals an instability when any odd $\tilde{f}_\ell$ approach $-1$, triggering PI in the corresponding channel. As we will see below the same is true in $g_{\ell}$'s, as predicted by Fermi liquid theory (Fig. \panelref{fig:fermisurfaces}{b}). In fact Eq. (\ref{eq:S_1}) is more suitable for detailed study of PI in even channels; for odd channel PIs it is better to use the dual version of (\ref{eq:S_1}). To be specific we will consider PI in the $\ell = 2$ channel below, with $g_2 \rightarrow 0^+$ and all other Landau parameters $g_{\ell \neq 2}, \tilde{g}_\ell > 0$. 

The (now $\mathbf{q}$-independent) coupling matrix $M_{\ell \ell'}'$ has, when
$g_2 = 0$, two exact nontrivial eigenvectors of zero eigenvalues;
$u^{(1)} = (\ldots, 0, - 1, 1, 0, \ldots)$ with $u^{(1)}_{1}=-u^{(1)}_{-1}=1$, and
$u^{(2)} = (\ldots, - 1, 1, 0, 0, 1 - 1, \ldots)$ with
$u^{(2)}_{1}=u^{(2)}_{-1}=0$ (Appendix \ref{sec:zero}).  This renders
$\Phi = i (\tilde{\varphi}'_1 - \tilde{\varphi}'_{- 1}) / \sqrt{2}$ and
$\Psi = (\tilde{\varphi}_3' + \tilde{\varphi}_{- 3}') - (\tilde{\varphi}_5' + \tilde{\varphi}_{- 5}') +
(\tilde{\varphi}_7' + \tilde{\varphi}_{- 7}') - \cdots$ as two potential critical fields that are real (instead of complex). 
$\Psi$ is not normalizable and has vanishing overlap with $\rho_2$. Furthermore, when $g_2$ turns negative, $\Phi$ becomes an eigenstate with eigenvalue $g_2=-|g_2|$, while the corresponding eigenvalue of $\Psi$ is $\sim -g_2^2$ and therefore much higher for small $|g_2|$ (Appendix \ref{sec:zero}). Upon projection to the critical subspace (details in
Ref. \cite{LongPaper}), we have:
\begin{equation}
    \rho (\theta, \mathbf{q}) = | \mathbf{q} | \sin (2 (\theta -
\theta_{\mathbf{q}})) \Phi (\mathbf{q}) / \sqrt{2}
    \label{eq:Densities}
\end{equation}
or $\rho_2 (\quu) = - \frac{i}{2\sqrt{2}} | \quu | e^{- 2 i \theta_{\quu}}
  \Phi (\quu)$.
Thus, $\Phi$ is the critical field and $\Psi$ can be safely ignored in the critical theory.

\begin{figure*}[htb]
  \centering
  \resizebox{\linewidth}{!}{
    \includegraphics{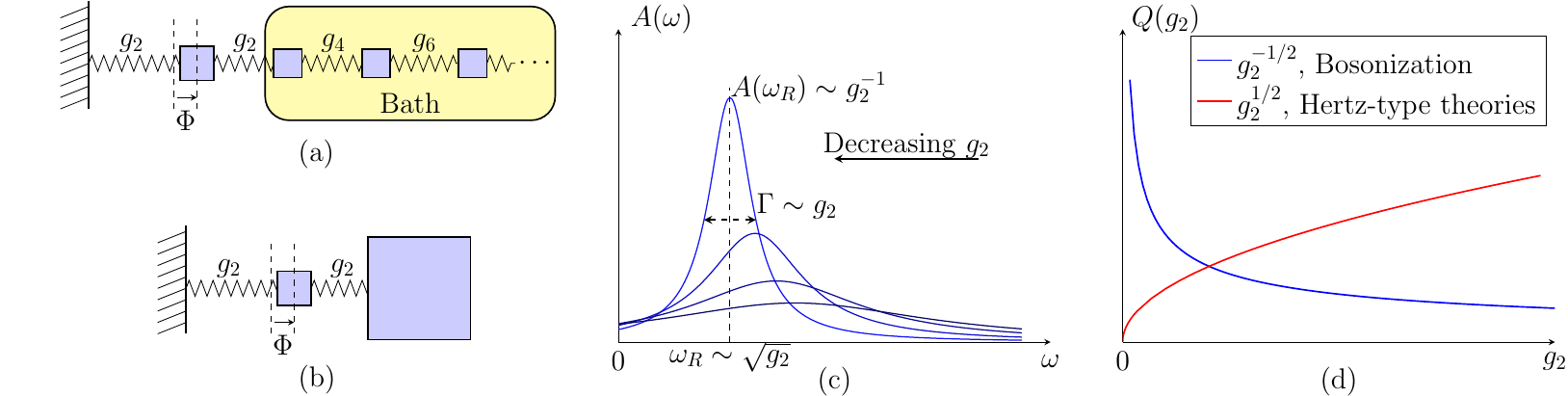}
  }
  \caption{Schematic representations of Fermi liquid modes close to the Pomeranchuk instability. (a) After scaling out the $|\quu|^2$ factor, the Hamiltonian  $M'_{\ell\ell'}$ resembles a soft oscillator $\Phi$ weakly coupled to an infinite transmission line. (b) Near criticality ($g_2\rightarrow 0$), the oscillators in the bath are coupled infinitely strongly in units of $g_2$, thus effectively forming a single block with infinite mass. (c) The spectral function $A(\omega)$ for the soft mode for different $g_2$ and (d) quality factors $Q(g_2)$ of our theory versus Hertz-type theories.}
  \label{fig:OscillatorsFigures}
\end{figure*}

Unlike the Hertz-Millis approach, the critical theory derived from bosonization does \textit{not} feature a Landau-damped critical mode. Close to criticality $(g_2 \rightarrow 0^+, g_{\ell \ne 2} \gg g_2)$, the matrix $M'_{\ell\ell'}$ can be mapped to a classical Caldeira-Leggett type system \cite{caldeira1983quantum} where the critical mode $\Phi$ is coupled to an infinite bath of harmonic oscillators via the coupling $g_2$. The spring constant of $\Phi$ is also $g_2$ (illustrated in Fig. \panelref{fig:OscillatorsFigures}{a}). 
Intuitively, the critical oscillator is damped by the infinite bath, but the damping rate would be suppressed by a factor of $g_2^2$. This impression is confirmed in the calculation of effective action of the critical oscillator (see below), which renders damping rate $\Gamma \sim g_2$ and quality factor:
\begin{equation}
    Q[\Phi] \equiv \frac{\omega_R}{\Gamma} \sim g_2^{-1/2}.
    \label{eq:q-factor}
\end{equation}
where $\omega_R \sim \sqrt{g_2}$ is the resonance frequency of critical mode. Thus, the critical mode is underdamped, and becomes sharper and sharper as we approach criticality (Fig. \panelref{fig:OscillatorsFigures}{c}). 
This is because as $g_2\rightarrow 0$, at the scale of $\omega_R$ the remaining blocks are effectively all locked together (because their couplings are much stronger than $g_2$) and form a single infinitely massive block behaving just like a fixed wall (see Fig. \panelref{fig:OscillatorsFigures}{b}), rendering the $\Phi$ oscillator undamped.

To illustrate this important point more explicitly and quantitatively, we set
all other Landau parameters to zero $(g_{\ell \ne 2} = \tilde{g}_\ell = 1)$, in which
case exact results can be obtained. Integrating out the non-critical modes {\em
  exactly} in the Gaussian action, we get the effective Lagrangian for $\Phi$ in
Euclidean signature as \cite{LongPaper} (see Appendix A for a derivation
and sanity check using free-fermion density-density correlation)
\begin{equation}
  \mathcal{L}^{(E)}_{\Phi} =
  \Bigg[\frac{\omega^2}{v_F} + \frac{g_2 v_F |\quu|^2}{4} - \frac{g_2^2}{a(s)+g_2} \frac{v_F |\quu|^2}{4}\Bigg] \Phi^*\Phi
  \label{eq:Leff}
\end{equation}
where $a (s) = 2 s  \left( s + \sqrt{1 + s^2} \right), s=|\omega|/(v_F |\quu|)$. Thus we have imaginary part of retarded self-energy being $\mathrm{Im} \Sigma_R(\omega_R) \sim g_2^{3/2}$, which gives the previous damping rate $\Gamma = -\mathrm{Im} \Sigma_R(\omega_R) / \omega_R \sim g_2^{}$ and quality factor $Q \sim g_2^{-1/2}$ (eq.~\eqref{eq:q-factor}).
Expanding Eq. \eqref{eq:Leff} in powers of $g_2/a(s)$ and keep to second order of $g_2$, we get
\begin{equation}
  \mathcal{L}^{(E)}_{\Phi} \approx
  \left[\frac{\omega^2}{ v_F} + \left(g_2 + \frac{g_2^2}{2} - g_2^2\frac{\sqrt{1+s^2}}{2 s}\right)\frac{v_F |\quu|^2}{4}\right] \Phi^*\Phi.
  \label{eq:Leff1}
\end{equation}
Again, we see that in this regime, the real part of the self-energy yields $\omega \sim \sqrt{g_2} v_F |\quu|$ dispersion and the leading order damping term is $g_2^2/s \sim g_2^{3/2}$, which is subleading to the dispersion term \footnote{There is never a lower order term in perturbative expansion of $g_2$ since $s$ appears only as powers of $g_2 / a(s)$}. The critical mode is underdamped in a finite window close to criticality.
On the other hand, one can set $g_2=1$ in the last term of Eq. \eqref{eq:Leff} and take the lowest order of the damping term (which is $s$) to obtain a Hertz-Millis-type theory.

Generalization to the case where PI occurs in an arbitrary $\ell=2n$ channel, $g_{2n}= 0$, is straightforward. Similar to the $\ell = 2$ case, the Fourier symbols for the eigenvectors of zero eigenvalues are $u^{(1)}(\theta)=\sin(\theta)-\sin(3\theta)+\cdots+(-1)^{n+1} \sin((2n-1)\theta)$
and $u^{(2)}(\theta)=\cos((2n+1)\theta)-\cos((2n+3)\theta)+\cdots$, whose Fourier transformations give the vectors $u^{(1)}$ and $u^{(2)}$ (Appendix \ref{sec:zero}). Consequently, we get $\rho(\theta) \propto u^{(1)}(\theta) \cos(\theta)\sim\sin(2n\theta)$, which means that the instability occurs in the angular momentum $2n$ channel, as expected.

\textit{Critical theory.}
For the critical mode, we need to introduce the next order term in the gradient expansion to stabilize the action \cite{Kun2004, Kun2005}. This term arises from interactions of the kind $\Dee\tvphi(\theta) \nabla^2 \Dpr \tvphi(\theta')$ ($|\nabla\rho|^2$ in terms of densities). 
Apart from the $\quu^4$ prefactor, such a term features exactly the same coupling
structure as the $\quu^{2}$ term, resulting in the
following effective Gaussian action for the Fermi liquid near a PI in an even
channel $2n$ (we rescale $\tau,\mathbf{x},\Phi$ to make $v_F=1$ and $\nabla^2 \Phi$ has coefficient $1$):
\begin{equation}
    S_0= \int\frac{d^2\quu d\omega}{(2\pi)^2} [\omega^2 - (1+f_{2n})|\quu|^2 - |\quu|^4]\Phi^\dagger(\quu)\Phi(\quu),
    \label{eq:actionOrdered}
\end{equation}
or in a real space form,
\begin{multline}
    \label{eq:CriticalTheory}
    S_0 = \int d^2 \mathbf{x}d t\  [\dot{\Phi}^2 - (1 + f_{2n}) |\nabla \Phi|^2
    - (\nabla^2\Phi)^2].
\end{multline}
This action is the 2D generalization of that in Ref. \cite{Kun2004}, and has dynamical exponent $z = 2$ at the PI $f_{2n} = -1$. It is also known as the quantum Lifshitz model in field-theory description of Lifshitz points \cite{Chaikin_1995,Ardonne_2004}. Our main result is that it is the proper critical theory describing the 2D Pomeranchuk transition in any even channel $\ell = 2n$. 

Setting $\ell=2$ for simplicity, Eq.~\eqref{eq:CriticalTheory} with suppressed time dependence takes the form of the quadratic part of Ginzburg-Landau theory, in which $\Phi(\mathbf{x})$ is related to the local nematic order parameter via Eq.~\eqref{eq:Densities}. More precisely, the magnitude and phase of $\rho_2$ are given by $2\sqrt2|\nabla\Phi|$ and $2\theta_{\nabla\Phi}$ respectively. The orientation of the nematic order, $\hat{\textbf{n}}$ (see Fig. \panelref{fig:fermisurfaces}{b}) is given by $\hat{\textbf{n}} \sim \widehat{\nabla\Phi}$ \footnote{This is schematic because Eq.~\eqref{eq:Densities} gives a nonlocal relation between $\hat{\textbf{n}}$ and $\nabla\Phi$.}. 
Moreover, $\hat{\textbf{n}}$ and $-\hat{\textbf{n}}$ are identified with the \textit{same} order parameter configuration, as the nematic order is a headless director (the same statement is true for all even-channel densities).
This does not change our critical theory but will determine the properties of the ordered state. Finally, the transformation $\Phi(\textbf{x}) \rightarrow -\Phi(\textbf{x}),\rho_2(\mathbf{x}) \rightarrow -\rho_2(\mathbf{x})$ rotates the nematic orientation globally by $\pi/2$ ($\pi/2n$ for PI in $2n$-channel), which is a particle-hole transformation in the density channel that features the instability.

\textit{Higher-order terms and RG analysis.} We first note that any coupling between the non-critical $z = 1$ Fermi liquid modes and the critical field $\Phi$ beyond the quadratic $|\quu|^2$ level is irrelevant, as under $z = 2$ RG scaling these modes become infinitely stiff (or equivalently, the Fermi velocity flows to infinity) \cite{Kun2005, LongPaper}. This includes couplings which are generated at the quadratic level but at higher-order in the gradient expansion (beyond the $|\quu|^4$ term considered above). Thus, we only need to concern ourself with $SO(2)$ symmetric self-interaction terms of $\Phi$. In what follows, we analyze the stability of the Gaussian fixed point (Eq.~\eqref{eq:CriticalTheory}) in the presence of all such terms.

At the PI fixed point, the order parameter $\Phi$ has engineering dimension $[\Phi] = 0$. Let us consider the lowest dimension terms of $\Phi$'s $n$-th power, $O_{n} \sim \lambda_n|\nabla \Phi|^{n}$ (powers of $\Phi$ are not allowed since a constant shift of $\Phi$ is a change of gauge). These terms can be generated from nonlinearity of the fermion dispersion and possible higher order interactions $\int_{\theta_1,\ldots,\theta_n} V(\theta_1,\ldots,\theta_n) \prod_{i=1}^{n} (\mathcal{D}_{\theta_i} \tvphi )^{n}$. Under $z=2$ scaling $(\mathbf{x}' = \mathbf{x}/\zeta, t'=t/\zeta^z)$, $\lambda_n$ scales as $\lambda_n' = \lambda_n \zeta^{n-4}$, thus $O_n$ is irrelevant for $n > 4$ and marginal for $n=4$.

Self-interaction terms $O_n$ of odd $n$ are not allowed in the effective theory because they are non-analytic in gradient expansion.
Thus we only need to include the marginal term $O_4$ in the critical theory \footnote{The only other term that could be marginal is $(\nabla^2 \Phi)(\nabla \Phi)^2$, however it is forbidden because it is not invariant under the $\Phi(\textbf{x}) \rightarrow -\Phi(\textbf{x})$ global rotation of the nematic order parameter, which is a particle-hole transformation.} and arrive at the Ginzburg-Landau action 
\begin{equation}
    S= S_0 - \lambda_4 \int d^2 \mathbf{x} d t\ |\nabla \Phi|^{4}
    \label{eq:fullCriticalTheory}
\end{equation}
where $\nabla \Phi$ plays the role of Landau's magnetization. The one-loop correction to the coefficient $\lambda_4$ is given by
\begin{equation}
    \lambda_4(\zeta) = \lambda_4 - 36\lambda_4^2\ \int_{\Lambda/\zeta}^\Lambda \frac{d^2q}{(2\pi)^2}\int_{-\infty}^\infty d\omega \frac{|\quu|^4}{(\omega^2 + |\quu|^4)^2}
\end{equation}
Thus,  $\lambda_4$ is marginally \textit{irrelevant} and we find the $\beta$-function
\begin{equation}
    \beta(\lambda_4) = -9 \lambda_4^2 + \mathcal{O}(\lambda_4^6).
\end{equation}

Finally, let us comment on the time-dependent higher-order terms constructed in
Ref.~\cite{Delacretaz2022}. Firstly higher order terms with a single time derivative shift the phase space structure of the action in Eq.~\eqref{eq:varphiAction}, and the configuration-space Lagrangian needs to be re-derived from the beginning. On the other hand, if the subleading dynamical terms were to be manually projected to our critical theory, they will not generate any relevant self-interaction terms in $\Phi$ \footnote{There is a marginal term that might be generated from RG flow, $\sim \dot{\Phi}(\nabla\Phi)^2$ which is again disallowed because the term is odd under $\Phi(\textbf{x},t) \rightarrow -\Phi(\textbf{x},t)$}. 
This completes our proof of the fact that the Pomeranchuk transition in any even
channel $\ell \geq 2$ is governed by a Gaussian fixed point with $z = 2$. The theory
behaves as if it is at its upper critical dimension; the correlators are
mean-field like with logarithmic corrections (namely $\beta=1/2, \gamma=1, \nu=1/2, \eta=0,\alpha=0$, as a result of which the hyper scaling relations $\gamma + 2\beta = (d+z) \nu$, $\beta = (d+z-2+\eta) \nu / 2$ and $\alpha = 2 - (d+z)\nu$ are satisfied for $d=z=2$), just like {\em classical} $\phi^4$
theory in 4 dimensions \cite{Chaikin_1995}. Such log corrections from the
marginally irrelevant terms are known to exist in the gapless Luttinger liquid
in 1d, see \cite{Lukyanov_1998}, while in 2d from the stable FL side they are
linear in one-loop RG \cite{Chitov_1998}.

\textit{Summary and Outlook.} In this work, we further developed the framework of high-dimensional bosonization and used it to study Pomeranchuk transitions in 2D. 
The critical theory turns out to be a $z=2$ Gaussian theory for a real scalar field (Eq.~\eqref{eq:CriticalTheory}) at its upper critical dimension, with only one marginally irrelevant perturbation. We thus expect mean-field critical properties, with logarithmic corrections.
This is possible because Fermi liquid interactions, which drive the Pomeranchuk transition, enter as quadratic terms in the bosonized action, which can (and must) be treated {\em exactly}, on equal footing with the free fermion terms. In fact, hints of our central result, namely $z=2$, already exist for closely related transitions \cite{Schattner2016, Berg2019,Liu2022}, for which our general arguments are expected to apply. Indeed, it would be useful to derive the critical theory for the Ising nematic transition using the formulation of 2D bosonization as presented in this letter. We leave this line of inquiry to future works. Besides the $z=2$ critical mode, there is also a companion $z=1$ propagating mode of velocity $v_F/\sqrt{2}$, as we discussed in Appendix.~\ref{sec:rho2-propagator}, which might also be experimentally probed.

In an earlier work (Ref. \cite{Lawler2006}) on the the critical theory for Pomeranchuk transitions also based on bosonization, all density $\ell \neq 2$ modes are integrated out following a Hubbard–Stratonovich transformation on the free-fermion term. Expanding the $\langle \rho_{-2}(-\quu)\rho_2(\quu)\rangle$ propagator close to the instability, the authors find contributions from $z = 2$ and $z = 3$ poles and conclude that the critical point is governed by $z = 3$ scaling. A similar conclusion was reached by Ref. \cite{Nilsson_2005}. As we show in Appendix \ref{sec:rho2-propagator}, the $z = 3$ contribution comes from non-critical part of $\rho_2$. The $z = 2$ mode we discuss in this Letter is the only critical field in the theory. Curiously, Ref. \cite{Lawler2006} also discusses a (finite velocity) $z = 1$ pole that is sharp, but sits \textit{inside} the particle-hole continuum. We find this second decoupled mode as well (Appendix \ref{sec:rho2-propagator}), and the reason it remains sharp at the instability is similar to the critical $\Phi$ field. 

Subsequent work using bosonization has considered a critical boson coupled to (otherwise) free Fermi gas \cite{Delacretaz2022, Cai2025}, where they also find $z = 3$. This is not inconsistent with our results; such works consider a $z = 1$ critical boson coupled to the fermions via a constant Yukawa-type coupling, whereas in our critical theory a $z = 2$ mode gets \textit{decoupled} from the particle-hole continuum at criticality (Fig. \panelref{fig:OscillatorsFigures}{a}).

In the accompanying work \cite{LongPaper} we will flesh out the details of the full Gaussian action of the $\tvphi$ modes and show that it reproduces the known free-fermion results, including the zero and higher sound modes. More importantly, we will study the single fermion correlator at the critical point, where we expect non-Fermi liquid behavior (which is the original motivation of high-dimensional bosonization but with very limited success so far). Finally, we will consider the ordered phase and discuss its properties including Goldstone modes and topological defects.

\textit{Acknowledgments.} J.P. acknowledges the support from Florida State University through the FSU Quantum Postdoctoral Fellowship. This research is supported by the National Science Foundation Grant No. DMR-2315954. Most of this work was performed at the National High Magnetic Field Laboratory, which is supported by National Science Foundation Cooperative Agreement No. DMR-2128556, and the State of Florida.

\bibliography{bibliography}\clearpage

\onecolumngrid
\section*{End Matter}
\appendix
\section{Zero modes of the Hamiltonian at the Pomeranchuk instability}
\label{sec:zero}

The coupling matrix that appeared in Eq. \eqref{eq:action_1},
\begin{equation}
    M_{\ell \ell'} (\mathbf{q}) = \int_0^{\pi} \frac{d \theta}{\pi} e^{i (\ell' - \ell) \theta} \cos^2  (\theta - \theta_{\mathbf{q}}) + 
    \int_0^{\pi} \frac{d \theta d \theta'}{\pi^2} 
    e^{i (\ell' \theta' - \ell \theta)} 
    \cos (\theta - \theta_{\mathbf{q}}) \cos (\theta' - \theta_{\mathbf{q}})
    f (\theta - \theta'),
    \label{eq:coupling-mat-0}
\end{equation}
after a basis change $\tvphi_\ell(\quu) = -ie^{-i\ell\theta_\quu}\tvphi'_\ell(\quu)$, which removes the phase factors on off-diagonals, is
\begin{equation}
  M' = \frac{1}{4} \begin{pmatrix}
    \ddots & \ddots  \\
    g_4 & g_4 + g_2 & g_2 & \\
      & g_2& g_2 + g_0 & g_0 & \\
      &  & g_0 & g_0 + g_2 & g_2 & \\
      &  &     & g_2 & g_2 + g_4  & g_4\\
      &  &     &      & \ddots & \ddots
  \end{pmatrix}.
  \label{eq:coupling-mat}
\end{equation}
where $g_\ell = 1 + f_\ell$ and $\tvphi_{-\ell}(-\quu)M_{\ell\ell'} \tvphi_{\ell'}(\quu) = \tvphi'_{-\ell}(-\quu)M'_{\ell\ell'} \tvphi'_{\ell'}(\quu)$.

When a given $g_{2n} = 0$, the Hilbert space fragments into two decoupled sectors; a finite-dimensional part consisting of the modes $(\tvphi'_{-(2n-1)}, \tvphi'_{-(2n-3)},\ldots, \tvphi'_{2n-1})$ and an infinite-dimensional part consisting of all the other modes. From the finite sector, we have the zero mode $u^{(1)}_\ell = (-1)^{\frac{\ell-1}{2}}, \ell \in \{-(2n-1), -(2n-3), \ldots, 2n-1\}$. It is easy to verify that this is an exact zero mode:
\begin{equation}
    M'\cdot u^{(1)} = g_{2(n-1)} - g_{2(n-1)} + g_{2(n-1)} - (g_{2(n-1)} + g_{2(n-2)}) + g_{2(n-2)} + \ldots - g_{2(n-1)} = 0.
\end{equation}

After another basis change to
$\alpha_\ell \equiv \frac{\tvphi'_\ell(\textbf{q}) + \tvphi'_{-\ell}(\textbf{q})}{\sqrt{2}}, \beta_\ell \equiv
i \frac{\tvphi'_\ell(\textbf{q}) - \tvphi'_{-\ell}(\textbf{q})}{\sqrt{2}},\ell>0$, the
action \cref{eq:S_0} becomes
\begin{equation}
  \label{eq:Lag-alpha-beta}
  S_0 = \int d\omega d^2k \sum_{\ell\geq1,\ell\in\text{odd}}
  \left[\frac{\omega^2}{v_F \tilde{g}_{\ell}}(\alpha_{\ell}^{\ast} \alpha_{\ell} + \beta_{\ell}^{\ast} \beta_{\ell}) - \frac{v_F
    k^2}{4} (\alpha_{\ell}^{\ast} M^{\alpha}_{\ell \ell'} \alpha_{\ell'} + \beta_{\ell}^{\ast} M^{\beta}_{\ell \ell'} \beta_{\ell'})\right]
  = S_0^{\alpha} + S_{0}^{\beta}
\end{equation}
where $\alpha$'s and $\beta$'s look like semi-infinite chains with nearest neighbor couplings
\begin{equation}
  M^{\alpha} = \left(\begin{array}{ccccc}
    2g_0 + g_2 & g_2 &  &  & \\
    g_2 & g_4 + g_2 & g_{4} &  & \\
               & g_4 & g_{4}+g_{6} & \ddots\\
                &  & \ddots & \ddots
  \end{array}\right), \quad M^{\beta} = \left(\begin{array}{ccccc}
    g_2 & g_2 &  &  & \\
    g_2 & g_4 + g_2 & g_{4} &  & \\
        & g_4 & g_{4}+g_{6} & \ddots\\
        &   & \ddots & \ddots
  \end{array}\right).
\label{eq:M-alpha-beta}
\end{equation}
From this we can see that critical mode $\Phi$ for $g_2=0$ is nothing but $\beta_1$, while the other unnormalizable zero mode $\Psi$ is $\alpha_1 - \alpha_3 + \alpha_{5} - \cdots$.

With $g_{\ell \neq 2} = 1$ we can obtain an effective action for $\Phi$ by integrating
out the rest of the degrees of freedom on $\beta$-chain. To that end, we first perform a
fourier $\sin$ transform
$X (u) = \sqrt{\frac{4}{\pi}} \sum_{n \geqslant 5, \tmop{odd}} \beta_n \sin((n - 3) u)$ on
$\beta_5,\beta_7,\ldots$ to get the (Euclidean signature) action $S_0^{\beta,(E)}=\int d\omega d^2q \mathcal{L}^{\beta}$

\begin{multline}
  \frac{ \mathcal{L}^{\beta} }{v_F q^{2}} = \left( s^2 + \frac{g_2}{4} \right) \beta_1^{\ast} \beta_1 +
  \left( s^2 + \frac{1 + g_2}{4} \right) \beta_3^{\ast} \beta_3 \\+
  \frac{g_2}{4}  (\beta_1^{\ast} \beta_3 + c.c.) + \int_0^{\pi / 2} d u \left[ \left( \frac{s^2}{v_F^2} + \cos^2 u \right) X^{\ast} X + c(u)  (X^{\ast} \beta_3 + c.c.) \right]
\end{multline}
where $c (u) = \frac{1}{4} \sqrt{\frac{4}{\pi}} \sin (2 u), s^2 =
\omega^2/k^2$. Standard gaussian integration over $X(u)$ then over $\beta_{3}$ gives rise
to \cref{eq:Leff}. A sanity check on \cref{eq:Leff} can be obtained as
follows. Observing that total fermion density is
$\rho_0(\quu) \sim |\quu| \alpha_1(\quu)$ and that the $\alpha$-chain only differs from
$\beta$-chain by an $\alpha_1$ onsite term (\cref{eq:M-alpha-beta}), one immediately has
the $\alpha_1$ correlator
\begin{equation}
  \langle \alpha_1 (-\quu) \alpha_1 (\quu) \rangle = \frac{\omega^2}{v_F} + \frac{(2+g_2) v_F |\quu|^2}{4} - \frac{g_2^2}{2 s  \left( s + \sqrt{1 + s^2} \right) + g_2} \frac{v_F |\quu|^2}{4} \mathop{=}\limits^{g_2=1} \frac{v_F q^2}{2}[1+s(s+\sqrt{s^2+1})]
  \label{eq:Leff-alpha1}
\end{equation}
where $s=|\omega|/(v_F |\quu|)$. And so
$\langle \rho_0 (-\quu) \rho_0 (\quu) \rangle = \frac{q^2}{2} \langle \alpha_1 (-\quu) \alpha_1 (\quu) \rangle =
\frac{1}{v_F} \left( 1 - \frac{s}{\sqrt{1 + s^2}} \right)$ yields the free
fermion density-density two points function.

On the other hand, with $g_{\ell \neq 2} = 1$ in \cref{eq:M-alpha-beta} we can solve the following eigenvalue equations
\begin{equation}
    M^{i} u^{i} = -2 \lambda^{i} u^{i}, \quad i=\alpha,\beta
\end{equation}
for positive $\lambda^\alpha,\lambda^\beta$ and normalizable solutions $u^{i} = (c^{i}_1, c^{i}_2, \ldots )$. The solution goes as
\begin{equation}
  c^{i}_n = \left\{ \begin{array}{ll}
    c^{i}_1, & n = 1\\
    A_{i} (x_+(\lambda^i))^{n - 2}, & n \geqslant 2
\end{array} \right., \quad i=\alpha,\beta
\end{equation}
and $-1 <x_+(\lambda) \coloneq - 1 - \lambda + \sqrt{(\lambda + 1)^2 - 1} < 0$. They satisfy the equations
\begin{equation}
  (2 + 2 \lambda^\alpha + g_2) c^{\alpha}_1 + g_2 A_{\alpha} = 0, \qquad (2 \lambda^\beta + g_2) c^{\beta}_1 + g_2 A_{\beta} = 0          
\end{equation}
where $\lambda^\alpha,\lambda^\beta$ are related to $g_2$ by
\begin{equation}
  g_2 = - \frac{a(\lambda^\alpha) b(\lambda^\alpha)}{a(\lambda^\alpha) + b(\lambda^\alpha)}, \qquad g_2 = - \frac{a(\lambda^\beta) d(\lambda^\beta)}{a(\lambda^\beta) + d(\lambda^\beta)}
\label{eq:energies}
\end{equation}
with $a (\lambda) = 1 + 2 \lambda + x_+ (\lambda) = \lambda + \sqrt{(\lambda +
1)^2 - 1},b (\lambda) = 2 \lambda + 2$ and $d(\lambda) = 2 \lambda$. Expanding the right hand sides of Eq.~\eqref{eq:energies} to leading orders of $\lambda^\alpha,\lambda^\beta$ one finds 
\begin{equation}
  g_2 = - \sqrt{2 \lambda^\alpha} + O((\lambda^\alpha)^{3/2}), \qquad g_2 = - 2 \lambda^\beta + O((\lambda^\beta)^{3/2})
\end{equation}
Thus we conclude that when $g_2 < 0$, $\Phi$ evolves into a state of energy $g_2$ while $\Psi$ evolves into a state of energy $-g_2^2$ higher that $g_2$.

\section{Propagator of \texorpdfstring{$\langle \rho_{-2} \rho_2 \rangle$}{〈|ρ₂|²〉}}
\label{sec:rho2-propagator}

In this section, we discuss the propagator $\langle \rho_{- 2} (-\mathbf{q})
\rho_2 (\mathbf{q}) \rangle$, which includes contributions from modes with dynamical
exponents $z = 1, 2$ and $3$. As we will see, the $z = 2$ pole is our critical
field $\Phi$, the $z = 1$ pole is a companion propagating mode (there are $2 n - 1$ such modes for $\ell = 2 n$ channel Pomeranchuk transition), while the $z = 3$ pole is not a mode, but incoherent superpositions of particle-hole pairs, just as in
free Fermi gas.

Define real fields $\alpha_\ell \equiv \frac{\tvphi'_\ell(\textbf{q}) + \tvphi'_{-\ell}(\textbf{q})}{\sqrt{2}}, \beta_\ell \equiv i \frac{\tvphi'_\ell(\textbf{q}) - \tvphi'_{-\ell}(\textbf{q})}{\sqrt{2}},\ell\geq1$ in Appendix A.
In this basis, $\Phi = \beta_1$, $\Psi = \sum_{\ell \ge 3} (-1)^{(\ell-1)/2}  \alpha_\ell$, and $\rho_2$ is represented as (recall Eq.~\eqref{eq:Densities}, see also \cite{LongPaper})
\begin{equation}
  \label{eq:rho2} \rho_2 (\mathbf{q}) = \frac{| \mathbf{q} |}{2 \sqrt{2}} e^{-
  i 2 \theta_{\mathbf{q}}}  (\alpha_1 (\mathbf{q}) - i \beta_1 (\mathbf{q}) +
  \alpha_3 (\mathbf{q}) - i \beta_3 (\mathbf{q})).
\end{equation}
In the Gaussian theory, $\{\alpha_\ell\}$ are completely decoupled from $\{\beta_\ell\}$. Assuming $g_{l \neq 2} = 1$ and integrating out $\{ \alpha_{\ell}, \beta_{\ell}
\}_{\ell = 5, 7, \ldots}$, we have (in Euclidean signature) the respective inverse
propagators for $\alpha_1, \alpha_3$ and $\beta_1, \beta_3$:
\begin{equation}
  \label{eq:propagators} G_{\alpha}^{- 1} \sim q^2  \left( \begin{array}{cc}
    s^2 + 1 / 2 + g_2 / 4 & g_2 / 4\\
    g_2 / 4 & s^2 + 1 / 4 + g_2 / 4 - r (s) / 4
  \end{array} \right), \qquad G_{\beta}^{- 1} \sim q^2  \left(
  \begin{array}{cc}
    s^2 + g_2 / 4 & g_2 / 4\\
    g_2 / 4 & s^2 + 1 / 4 + g_2 / 4 - r (s) / 4
  \end{array} \right)
\end{equation}
where $r (s) = 1 + 2 s^2 - 2 s \sqrt{1 + s^2}, s = | \omega | / (v_F q)$.
Setting $g_2 = \delta + \kappa q^2, \delta \rightarrow 0^+$, $s^2 + \kappa
q^2$ in $G_{\beta}^{- 1}$ corresponds to our $z = 2$ critical mode and $s^2 +
1 / 2$ in $G_{\alpha}^{- 1}$ is a companion $z = 1$ mode. The $(2,2)$ elements of both $G_{\alpha}^{- 1}$ and $G_{\beta}^{- 1}$ stand for the effective actions of the rest of the two chains on their first sites. These effective actions mimic the behavior of free fermi gas density-density effective action which features Landau damping because of the coupling to the particle-hole continuum (an infinite bath).
Putting $g_2 = \delta + \kappa q^2, \delta
\rightarrow 0^+$, the term $(G_\alpha^{-1})_{22} = (G_\beta^{-1})_{22}$ is
\begin{equation}
  s^2 + 1 / 4 + \frac{\delta + \kappa q^2}{4} - r (s) / 4 = s^2 +
  \frac{\delta}{4} + \kappa q^2 - \frac{2 s^2 - 2 s \sqrt{1 + s^2}}{4}.
\end{equation}
Balancing $s \sqrt{1 + s^2} \sim s$ versus the $\kappa q^2$ term, we find $z = 3$
behavior, which is the claimed critical mode in Ref.~\cite{Lawler2006}.

\end{document}